\documentclass[prl,twocolumn,groupedaddress,nofootinbib,english,floatfix,superscriptaddress,preprintnumbers]{revtex4-2}

\usepackage{graphicx}
\usepackage{cancel}
\usepackage{amssymb}
\usepackage{textcomp}
\usepackage{amsmath}
\usepackage{mathtools}
\usepackage{bm}
\usepackage{times}
\usepackage{epsfig}
\usepackage[dvipsnames]{xcolor}
\usepackage{hyperref}
\usepackage{setspace}
\usepackage{comment}
\usepackage{subcaption}
\usepackage[utf8]{inputenc}
\usepackage[english]{babel}
\usepackage{textcomp}
\usepackage{latexsym}
\usepackage{mathrsfs}
\usepackage{flushend}
\usepackage[section]{placeins}
\usepackage[normalem]{ulem}
\usepackage{eso-pic}
\usepackage{pgffor}

\usepackage{tikz}
\usetikzlibrary{patterns}
\usetikzlibrary{shapes.geometric}
\usetikzlibrary{decorations.pathmorphing}
\usetikzlibrary{arrows.meta}
\tikzset{snake it/.style={decorate, decoration=snake}}
\usetikzlibrary{positioning}
\usetikzlibrary{arrows,shapes,positioning}
\usetikzlibrary{decorations.markings}
\tikzstyle arrowstyle=[scale=1]
\tikzstyle directed=[postaction={decorate,decoration={markings,mark=at position .65 with {\arrow[arrowstyle]{stealth}}}}]
\tikzstyle reverse directed=[postaction={decorate,decoration={markings,mark=at position .65 with {\arrowreversed[arrowstyle]{stealth};}}}]

\tikzset{->-/.style={decoration={
  markings,
  mark=at position #1 with {\arrow{>}}},postaction={decorate}}}

\tikzset{-<-/.style={decoration={
  markings,
  mark=at position #1 with {\arrow{<}}},postaction={decorate}}}

\DeclareMathAlphabet{\mathcalligra}{T1}{calligra}{m}{n}
\usepackage[utf8]{inputenc}

\newcommand{\subA}{\textrm{\tiny{A}}}
\newcommand{\subH}{\textrm{\tiny{H}}}

\newcommand{\subU}{\textrm{\tiny{U}}}

\newcommand{\csch}{\textrm{csch}}

\newcommand{\MCt}[1]{\textcolor{blue}{ #1}}

\begin{document}
\title{Massive quantum divergence on the Cauchy horizon of a 
black hole}
\author{Marc Casals}
\email{marc.casals@uni-leipzig.de}
\affiliation{Institut f\"ur Theoretische Physik, Universit\"at Leipzig,\\ Br\"uderstra{\ss}e 16, 04103 Leipzig, Germany}
\affiliation{School of Mathematics and Statistics, University College Dublin, Belfield, Dublin 4, D04 V1W8, Ireland}
\affiliation{Centro Brasileiro de Pesquisas F\'isicas (CBPF), Rio de Janeiro, CEP 22290-180, Brazil}

\author{Lorenzo Pisani}
\email{lorenzo.pisani2@mail.dcu.ie}
\affiliation{Centre for Astrophysics and Relativity, School of Mathematical Sciences, Dublin City University, Glasnevin, Dublin 9, Ireland.}

\author{Peter Taylor}
\email{peter.taylor@dcu.ie}
\affiliation{Centre for Astrophysics and Relativity, School of Mathematical Sciences, Dublin City University, Glasnevin, Dublin 9, Ireland.}

\begin{abstract}
We investigate a quantum {\it massive} scalar field in the interior of a charged and spherically-symmetric (Reissner-Nordstr\"om) black hole.
We examine the behaviour for varying values of the black hole charge, field mass and coupling constant when the field is in two quantum states: Hartle-Hawking (representing a black hole in thermal equilibrium) and Unruh (representing a black hole evaporating via the emission of Hawking radiation). We show that  
 the vacuum polarization (in both states) generically 
{\it diverges} on the Cauchy  horizon, in stark contrast to what happens for massless fields; also,
the angular components of the quantum stress-energy tensor generically
{\it diverge} on the Cauchy  horizon and, in at least one of the two states, they do so more strongly than for massless fields. 
Furthermore, we calculate the 
energy  fluxes in Eddington-Finkelstein coordinates $\{u,v\}$.
We show that these fluxes in both states on the Cauchy horizon do not generically vanish. This implies, in particular,
that in regular, Kruskal coordinates $\{U,V\}$ the ingoing flux diverges 
like $V^{-2}$ on the Cauchy horizon (where $V=0$).
This divergence suggests that its backreaction via the semiclassical Einstein equations would yield a strong singularity, as opposed to its weaker, classical counterpart. 
Interestingly, there are exceptions, in which the energy fluxes in Eddington-Finkelstein coordinates   vanish: (i) in the extremal limit (where the black hole is maximally charged); (ii) certain fine-tuned regions of parameter space, where the  fluxes change sign.
\end{abstract}

\maketitle


\paragraph*{Introduction.}

The interiors of black holes (BHs) are fascinating regions, as they may harbor spacetime singularities or hypersurfaces (called Cauchy horizons - CHs) beyond which the  classical Einstein equations cease to be deterministic (the Cauchy value problem is no longer well-posed).
It is well-known that the presence of classical massless fields may lead to the formation of a weak~\cite{Tipler1977singularities}, null curvature singularity along the would-be regular CH of isolated BHs -- see~\cite{ori1992structure,DafermosLuk2017,Brady1998late,Ori1999oscillatory} for neutral and  rotating (Kerr) BHs and~\cite{1981PhLA...83..110H,Poisson1990internal,Ori1991inner,Brady1995black,Hod1998mass,Burko1997structure,dafermos2005interior} for  charged and spherically-symmetric (Reissner-Nordstr\"om, RN) BHs. When a classical scalar field in RN has a mass, then the singularity becomes slightly stronger (namely, it does not admit locally integrable Christoffel symbols), but it continues to be null and continuously-extendible~\cite{kehle2024strong,2018CMaPh.360..103V,1998PhRvD..58b4018H,PhysRevD.63.064032,PhysRevD.88.024054,2004PhRvD..70d4018B,PhysRevD.68.044013}.

Importantly, it has been shown in  recent years that quantum physics may actually play  
a 
significant
role in
the nature of the structures that lurk inside BHs.
We consider the semiclassical framework whereby quantum fields propagate on a 
classical background spacetime. The quantum fields 
induce quantum corrections on the classical metric via
the semiclassical Einstein equation, which is sourced by
the renormalized expectation value of the quantum stress-energy tensor (RSET) $\langle \hat{T}_{\mu\nu}\rangle_{\textrm{ren}}^\subA$ for the matter fields in some quantum state $|A\rangle$.
It has been found that the 
flux components of the RSET
for massless matter fields diverge on the CH more `strongly' 
(such that the semiclassical backreaction on a fixed background would yield a discontinuous metric)
 than the corresponding classical fluxes
-- see~\cite{Zilberman:2019buh,Lanir:2018vgb,2021PhRvD.104b4066Z,alberti2026quantumfluxeslanglehatphi2ranglenonminimally} for RN;~\cite{hollands2020quantum,
Klein2024long} for a RN BH immersed in a de Sitter Universe (RNdS);~\cite{2022PhRvL.129z1102Z,Zilberman:2024jns,McMaken:2024fvq} for  
Kerr;~\cite{PhysRevLett.132.121501} for a Kerr BH immersed in a de Sitter Universe.

In this Letter we aim to contribute to 
understanding BH interiors
by investigating the effects of
a {\it massive} quantum scalar field $\hat\Phi$ in the interior of a RN BH spacetime
with varying coupling parameter $\xi$ between the scalar field and
the gravitational field.
A massive scalar field may be viewed as the simplest model for massive matter.
Also, the only scalar field we know exists (namely, the Higgs boson) is massive (even if it is self-interacting, and so beyond our free
field setting).
Furthermore, theories of particle physics beyond the standard model commonly predict the existence of massive scalar fields (e.g.,~\cite{PhysRevLett.40.223,arvanitaki2010string,arvanitaki2011exploring}).

We consider two quantum states of physical interest: the Unruh state~\cite{unruh1976notes},
which describes the field around an astrophysical BH evaporating via the emission of Hawking radiation;
and the Hartle-Hawking (H-H) state~\cite{hartle1976path},
which describes the field in thermal equilibrium with the BH.
In particular, we find that the (renormalized) vacuum polarization (VP) 
$\langle \hat{\Phi}^{2}(x)\rangle_{\textrm{ren}}^{\subA}$ 
diverges for points $x$ on the CH where,
here and henceforth, we use $\subA$ to denote generically the Unruh (\subU) or H-H (\subH) state.
This divergence of the VP may come as a surprise, since the VP for massless fields has so far been seen to be regular on the CH in all settings investigated -- see~\cite{Lanir:2018vgb} in RN and~\cite{Zilberman:2024jns,alberti2026quantumfluxeslanglehatphi2ranglenonminimally} in polar Kerr.
Apart from the fact that $\langle \hat{\Phi}^{2}(x)\rangle_{\textrm{ren}}^{\subA}$ and $\Box \langle \hat{\Phi}^{2}(x)\rangle_{\textrm{ren}}^{\subA}$  yields  (up to geometrical terms) the trace of the RSET,
the VP is also interesting in its own right: 
it characterizes the
quantum fluctuations
of a quantum vacuum and, as such,
 it has been employed as a local definition of temperature in curved spacetimes \cite{Buchholz2007, Sanders2017} and  
 as a model of quantum fluctuations of lightcones and null horizons \cite{Ford1995,FordSvaiter1996, FordSvaiter1997}.
Furthermore, 
within 
interacting theories,
the VP provides information about spontaneous symmetry breaking near a BH~\cite{1982qsst.conf..131F,1981CMaPh..80..421H}.

We also calculate the RSET throughout the BH interior, with a particular focus on the CH.
Similarly to the VP, our results for the angular components of the RSET also diverge on the CH in the massive case, in  a manner which,  at least for one of the two states, is more strongly  than in the massless case; for the specific coupling $\xi=1/4$ they do not diverge for massless fields whereas we show that they do for massive ones.
With regards to the important energy fluxes,
we show that, in  Kruskal coordinates $\{U,V\}$, which are regular on the CH, where 
$U=0$ or
$V=0$, the  
outgoing and ingoing 
fluxes in both states diverge like,
respectively, $U^{-2}$ or
$V^{-2}$, except for certain fine-tuned parameters, namely, for maximally charged black holes and along a line in the field mass-coupling parameter space.
This quadratic divergence in the quantum setting is stronger  than the corresponding one in the classical setting.
Surprisingly, 
however,
this  
quadratic
divergence is the same as for massless fields~\cite{Zilberman:2019buh}, despite the fact that the VP diverges for massive fields but is regular in the massless case
and that, in the classical setting, the derivatives of the field (from which the classical stress-energy tensor is --partly-- obtained) are more irregular when including a mass.
On the other hand, 
the presence of the field mass does change the oscillatory behaviour of the ingoing flux  near maximal BH charge.
Lastly, in the Supplemental Material (SM) we provide plots of both the VP and the RSET for massive fields {\it throughout} the RN BH interior. These very accurate numerical evaluations of the RSET 
throughout the interior are essential to compute the backreaction on the interior via the semiclassical Einstein equations. 

We note that
Ref.~\cite{Hollands2020quantum2} already considered a massive scalar field on the CH of a BH, namely, of a RNdS BH.
However, there are many significant differences between Ref.~\cite{Hollands2020quantum2} and our work.
First, the calculation in Ref.~\cite{Hollands2020quantum2} is limited to the influx only  
on the CH, away from near-extremality and for zero coupling constant. On the other hand,
our results (in RN) are for the VP as well as all components of the RSET, everywhere throughout the interior, for arbitrary values of the field and spacetime parameters -- including arbitrarily close to extremality and arbitrary coupling constant. Second, RNdS 
contains a  redshift
effect from the cosmological expansion which has a substantial effect on the regularity properties of the CH. 
For example, classically there is violation of strong cosmic censorship in RNdS~\cite{Cardoso:2017soq} but not in RN~\cite{luk2017proof};
semiclassically,
we find (in later Figs.~\ref{fig:Tyy_CH_Mass} and \ref{fig:Tyy_CH_Q}) that the leading order divergence is state dependent 
whereas Ref.~\cite{Hollands2020quantum2} find that the divergence is agnostic to the choice of (Hadamard) state.

Possible reasons why a comprehensive analysis of quantum effects for a massive field on BH interiors had not so far been conducted may lie in: (i) the  
significant technical challenge
of renormalization, 
which is here
compounded by (ii)
the additional  
difficulty  
of dealing with `physical' logarithmic (see the later Eq.~\eqref{eq:asympt VP}) divergences (on the CH),  combined with (iii) the 
very surprising feature that one must 
 calculate renormalized quantities extraordinarily close 
to the CH in order to peel off their asymptotic behaviour
(see the later Figs.~\ref{fig:vacuum_polar_IH_different_masses_U} and \ref{fig:results_diffQ}), which presents a significant numerical challenge.
Our massive field results have been enabled by implementing
a renormalization technique called the
extended coordinate method~\cite{taylor2016mode,taylor2017mode,taylor2022mode,arrechea2025renormalized},
which
 involves point splitting in all directions except the radial  
 one.
The extended
coordinate method solves all previous problems, since it is agnostic to field parameters (like the mass, and therefore can handle the logarithmic divergences in the case of a massive field), and is extremely accurate and efficient (thus allowing for high numerical precision).
This method was originally developed 
for 
fields in the H-H state 
in the exterior of spherically-symmetric BHs,  where the Lorentzian metric can be Wick-rotated into a Riemannian metric. This Wick-rotation greatly facilitates the implementation of the  
method, since the Euclidean Green function has a simpler singular support (namely, only for coincident spacetime points, as opposed to along null geodesics in the Lorentzian case) and the continuous frequency-integral when decomposing into Fourier modes  becomes a discrete sum. Even though Wick-rotation is, unfortunately, no longer possible in the interior of a BH (where the metric is not static), we still managed to implement the extended coordinate method, here for the first time in  the case of a Lorentzian metric.

We choose units with $c=G=\hbar=1$.


\paragraph*{Massive quantum field in RN.}

\begin{figure}
\includegraphics
[width=7cm]
{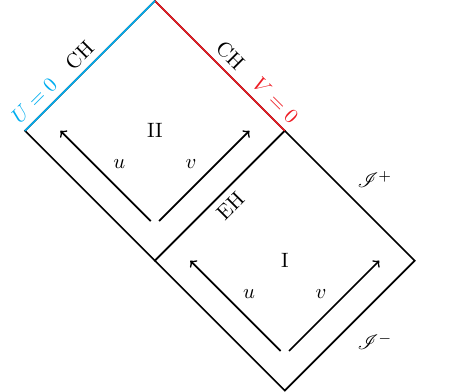}
\caption{\small Penrose diagram of (part of) RN spacetime.}
\label{fig:PenroseDiag}
\end{figure}

\begin{figure}
\includegraphics[width=\linewidth]{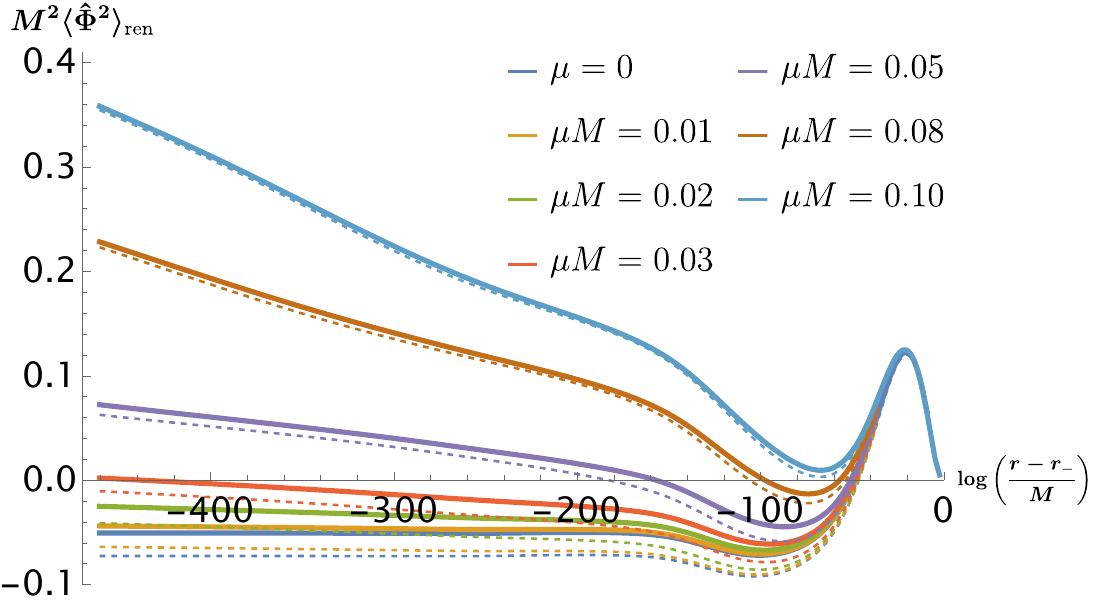}
\caption{\small VP $\langle\hat\Phi^2\rangle_{\mathrm{ren}}^\subA$ in the H-H (solid lines) and Unruh (dashed lines) states for BH charge $Q=0.8 M$ and $\ell=M$ near the CH for different values of the field mass $\mu$. 
}
\label{fig:vacuum_polar_IH_different_masses_U}
\end{figure}

\begin{figure}
\includegraphics[width=\linewidth]{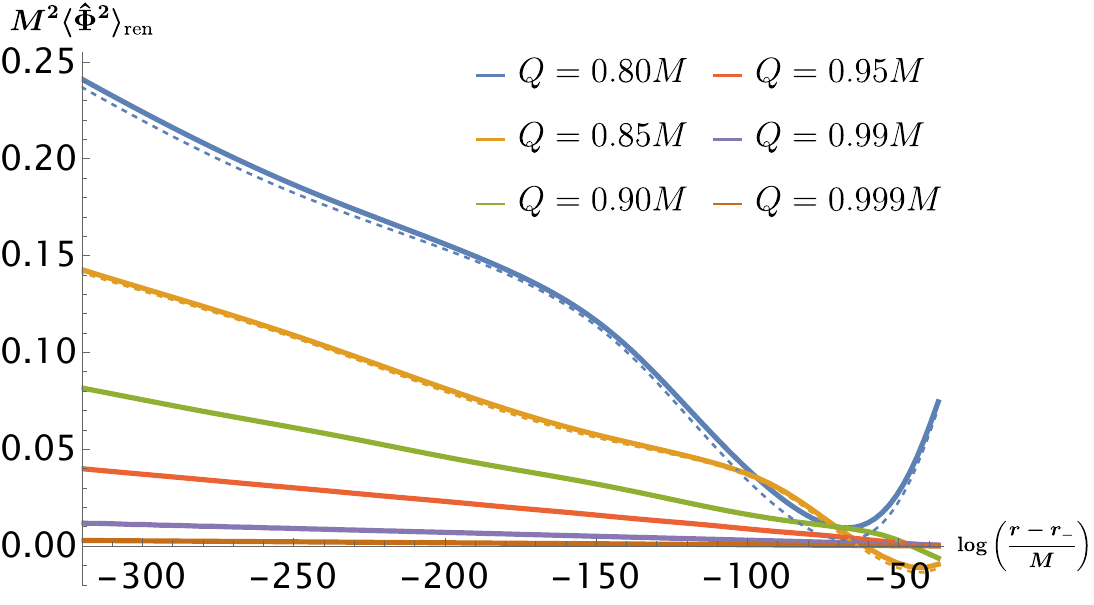}
\caption{\small  VP $\langle\hat\Phi^2\rangle_{\mathrm{ren}}^\subA$ in the H-H (solid lines) and Unruh (dashed lines) states for field mass $\mu M =0.1$ and $\ell=M$ near the CH for different values of the BH charge $Q$. 
}
\label{fig:results_diffQ}
\end{figure}

We consider the interior of a  
RN BH spacetime with mass $M$ and charge $Q<M$. Its metric can be written as
\begin{align}
\label{eq:metric}
    ds^{2}&=-f(r)dt^{2}+\frac{dr^{2}}{f(r)}+r^{2}\left(d\theta^{2}+\sin^{2}\theta\,d\phi^{2}\right),
\end{align}
in coordinates
$(t,r,\theta,\phi)\in \mathbb{R}\times (r_-,r_+)\times\mathbb{S}^2$, 
where 
$f(r):=
\frac{(r-r_+)(r-r_-)}{r^2}$ and
 the event horizon ($+$) and 
CH ($-$) 
have radii 
$r_{\pm}=M\pm \sqrt{M^2-Q^2}$ and surface gravities 
$\kappa_{\pm}=
\frac{r_\pm-r_\mp}{2r^2_{\pm}}\gtrless 0$.
Clearly, the maximal BH charge is $Q=M$, corresponding to an `extremal' BH.
It is also useful to define two sets of double null coordinates in the interior ($r<r_+$): Eddington-Finkelstein coordinates $u=r_*-t$, $v=t+r_*$ and Kruskal coordinates 
$U=-e^{\kappa_- u}$, $V=e^{\kappa_- v}$,
where   $r_*=r+\frac{1}{2\kappa_+}\ln\left|\frac{r-r_+}{r_+-r_-}\right|+\frac{1}{2\kappa_-}\ln\left|\frac{r-r_-}{r_+-r_-}\right|$.
The coordinates $\{U,V\}$ are regular across the event horizon whereas $\{u,v\}$ are not.
While for an eternal RN BH, the CH is formed by 
the ingoing section $V=0$ and the outgoing section $U=0$, 
for an astrophysical BH formed from gravitational collapse, the CH is only  
the
$V=0$ section, which is thus our main interest.
See Fig.~\ref{fig:PenroseDiag}.

The Klein-Gordon equation for a scalar field $\Phi$ is
\begin{align}
\label{eq:wave_eqn}
    (\Box-\mu^2-\xi\,R)\Phi=0,
\end{align}
where $\Box$ is the d'Alembertian operator, $\mu$ is the mass of the scalar field, $R$ is the Ricci scalar (which vanishes in RN) and $\xi$ is the coupling of the field to the background curvature.
The field $\Phi$ can be decomposed into 
spherical-harmonic (labelled by $l$) and Fourier (labelled by the frequency $\omega$)
modes
     $\Phi_{\omega l}=e^{-i\omega t}Y_{lm}(\theta,\phi)\psi_{\omega l}(r)/r$,
 where $\psi_{\omega l}$ satisfies
 \begin{align}
 \label{eq:Schrodinger}
 \left\{\frac{d^{2}}{dr_{*}^{2}}+
 \omega^{2}-f(r)\left(\mu^2+\frac{l(l+1)}{r^{2}}+\frac{f'(r)}{r}\right)
 \right\}\psi_{\omega l}(r)=0.
 \end{align}
 It is useful to define two solutions of \eqref{eq:Schrodinger}.
 In the BH exterior, we define $\psi^{\textrm{up}}_{\omega l}(r)$ via
\begin{align}
\label{eq:LorentzianBCUp}
\psi^{\textrm{up}}_{\omega l}(r)&\sim
\begin{cases}
    \displaystyle{e^{i\omega r_{*}}+A_{\omega l}^{\textrm{up}}e^{-i\omega r_{*}}},\qquad & r\to r_{+}\\ \\
    \displaystyle{B_{\omega l}^{\textrm{up}} e^{i\tilde{\omega}  r_{*}}},  & r\to\infty,
\end{cases}
\end{align}
where $\tilde{\omega}\coloneqq\textrm{sgn}(\omega)\,\sqrt{\omega^{2}-\mu^{2}}$
for $\omega^2>\mu^2$
and $\tilde{\omega}\coloneqq i\,\sqrt{\mu^2-\omega^2}$
for $\omega^2<\mu^2$,
for some exterior  coefficients $A_{\omega l}^{\textrm{up}}$ and $B_{\omega l}^{\textrm{up}}$.
We also define a solution $\psi_{\omega l}^{\textrm{II}}(r)$ in the BH interior  via
\begin{align}
\label{eq:phi_asympt}
  \psi_{\omega l}^{\textrm{II}}(r)
    &\sim
    \begin{cases}
    \displaystyle{ 
   A_{\omega l}e^{i\omega r_{*}}+ 
   B_{\omega l}e^{-i\omega r_{*}}
  }
 ,\qquad & r\to r_{-},\\ 
 \\
    \displaystyle{
   e^{-i\omega r_{*}}
    }
    , & r\to r_+,
\end{cases}
\end{align}
for some interior  coefficients $A_{\omega l}$ and $B_{\omega l}$.

We next quantize the field by expanding it in a basis of modes, promoting the coefficients and field to operators ($\Phi\to\hat\Phi$) and imposing canonical commutation relations. The choice of basis essentially determines the quantum state $\textrm{A}$ of the field.
The VP can be calculated as $
\langle \hat{\Phi}^{2}(x)\rangle_{\textrm{ren}}^{\subA}=\lim_{x'\to x} \left(G_\subA(x,x')-K(x,x')\right)$, where 
$G_{\subA}(x,x')=\langle \textrm{A} |\hat{\Phi}(x)\,\hat{\Phi}(x')|\textrm{A}\rangle$ is the Wightman function and 
$K(x,x')$ is the Hadamard parametrix 
(see 
Eq.~(5)
in SM).
This parametrix depends on 
an arbitrary lengthscale 
$\ell$ which is 
associated with the renormalization ambiguity~\cite{wald1994quantum}.

The states of physical interest, namely Unruh  
and H-H,
are typically defined using Kruskal coordinates but, unfortunately, the field does not admit separation in these coordinates.
One must thus re-express the Wightman function in the Unruh and H-H states in terms of separable modes $\Phi_{\omega l}$. We  
obtained
these expressions 
(see Eqs.~(2) and (4)
of the SM)
for 
the Wightman function for
a massive field in the interior of RN 
for the first time, to the best of our knowledge.
As for the Hadamard parametrix, we first expand it for small distances with $r=r'$ (see 
Eq.~(7)
in SM) and then, within our extended coordinate method, we 
decompose the expansion into spherical-harmonic and Fourier modes.
We put together the mentioned 
mode-decompositions
 for $G_{\subA}$ and $K$ to obtain  
 an expression for 
$\langle \hat{\Phi}^{2}(x)\rangle_{\textrm{ren}}^{\subA}$ (Eq.~(8)
 in SM),
which 
is at the basis of our calculations
in the next section.


\paragraph*{Results for VP.}

Whereas away from the CH, the general behaviour of the VP (in both the Unruh and H-H states) is similar for massless and massive fields  -- see  
Fig.~1
in SM --, their behaviour is fundamentally different in their approach to the CH,
as we  show below both numerically and semianalytically.
In the massless case, when the VP   approaches the CH,  after a few oscillations, it asymptotes to a {\it finite} 
value.
In the massive field case, 
on the other hand,
the VP  (in both the Unruh and H-H states) actually (logarithmically) {\it diverges} at the CH.

We first show this divergence numerically in Figs.~\ref{fig:vacuum_polar_IH_different_masses_U} and \ref{fig:results_diffQ} for different field masses and BH charges
(see 
Sec.~IV
in SM for details on the numerical evaluation of $\langle \hat{\Phi}^{2}(x)\rangle_{\textrm{ren}}^{\subA}$).
Both figures show that the divergence is of the form  
\begin{equation}\label{eq:asympt VP}
\langle \hat{\Phi}^{2}(x)\rangle_{\textrm{ren}}^{\subH/\subU}\sim k(\mu,Q)\log|f|, \qquad r\to r_-,
\end{equation}
with $\lim_{\mu\to 0}k(\mu,Q)=0$.
Fig.~\ref{fig:vacuum_polar_IH_different_masses_U} further shows that the VP approaches the massless (finite) result in the  limit  $\mu\to 0$. 
It also shows a suppression of the state dependence for larger values of field mass. 
 Fig.~\ref{fig:results_diffQ} further shows that,  as the BH charge approaches the extremal value $Q=M$, the onset of the asymptotic logarithmic behaviour lies further away from the CH, while the coefficient $k(\mu,Q)$  becomes smaller in such a way that $\lim_{Q\to M}k(\mu,Q)=0$.
We also note that the behaviour is rather similar between the Unruh and H-H states.

We were also able to derive {\it semi-analytically} the logarithmic divergence of the massive VP in the CH limit.
 The result (see Sec.~IV in SM) is
 Eq.~\eqref{eq:asympt VP} with
\begin{align}
\label{eq:nearhorizonclosed}
     k(\mu,Q)=
     \sum_{l=0}^{\infty}(2l+1)h_l(\mu),
\end{align}
where
\begin{equation}\label{eq:hl}
    h_l(\mu):=
    \frac{1}{16\pi^{2}r_{-}^{2}}
    \left(
    1-\frac{4 r_{+}^{2}r_{-}^{2}}{(r_{+}-r_{-})^{2}}\left(\pi^{2}q_{l}^{2}+\kappa_{+}^{2}
    c_{l}^{2}
    \right)
    \right),
    \end{equation}
\begin{align}
    c_{l}
    := \Re\{a^{(0)}_{l}\}-\Re\{b^{(0)}_{l}\}+q_{l}\,\Im\{A^{(1)}_{l}\}\,\in\mathbb{R},
\end{align}
and 
where we have made use of
the small-$\omega$  expansion~\cite{McNamara1978, Kehle2019,RSETLong} of the 
scattering coefficients: 
\begin{align}\label{eq:small-w Coeffs}
A_{\omega l}
    &=\frac{i\,q_{l}}{\omega}+a^{(0)}_{l}+\mathcal{O}(\omega),\quad
    B_{\omega l}
    =-\frac{i\,q_{l}}{\omega}+b^{(0)}_{l}+\mathcal{O}(\omega),\nonumber\\
    A^{\textrm{up}}_{\omega l}&=-1+
   A_{l}^{(1)}\,\omega+
   \mathcal{O}(\omega^{2}),
\end{align}
where $q_{l},
iA^{(1)}_{l}\in \mathbb{R}$, 
$a^{(0)}_{l},b^{(0)}_{l}\in \mathbb{C}$,
$\Im\{b^{(0)}_{l}\}=-\Im\{a^{(0)}_{l}\}$.

\begin{figure}
\centering
\includegraphics[width=\linewidth]{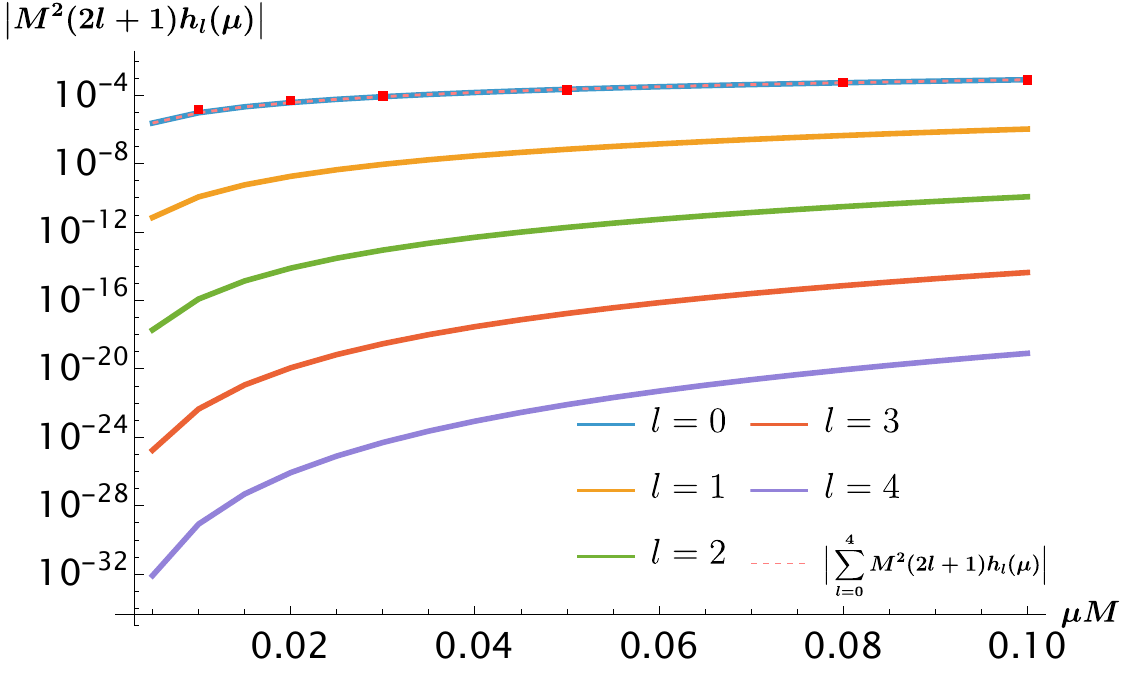}
\caption{
Log-plot of 
$k(\mu,Q)$ 
in \eqref{eq:asympt VP}
as well as its various $l$-mode contributions as per 
\eqref{eq:nearhorizonclosed} 
as a function of the field mass $\mu$ with BH charge $Q=0.8M$. 
The red squares represent  
$k(\mu,Q)$ 
obtained by fitting the numerical results for  $\langle\hat\Phi^2\rangle_{\textrm{ren}}^\subH$.}
\label{fig:log_coeff_num}
\end{figure}

\begin{figure}
\centering
\includegraphics[width=0.9\linewidth]{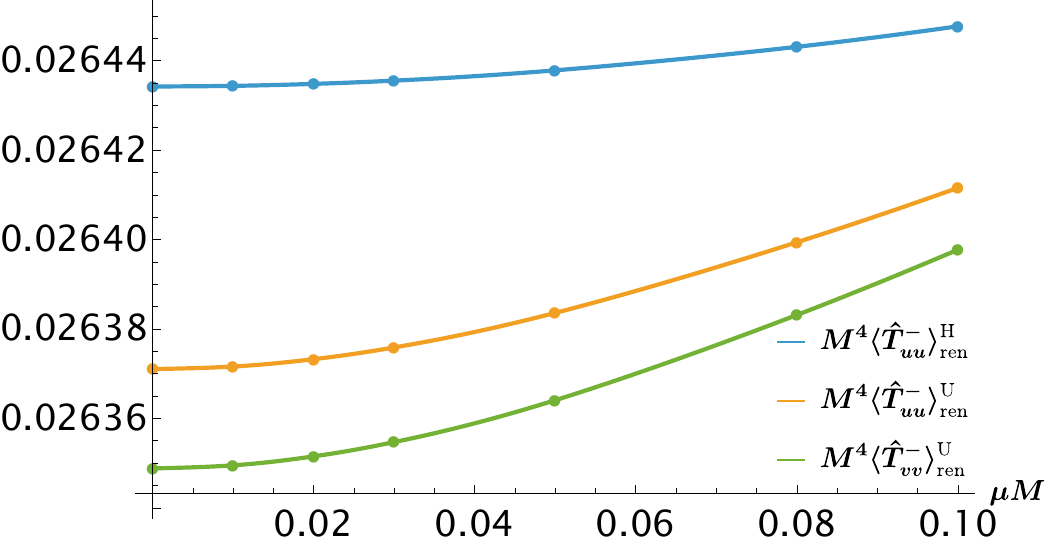}
\caption{CH fluxes $\langle\hat{T}_{uu}^-\rangle^\subH_\textrm{ren}=\langle\hat{T}_{vv}^-\rangle^\subH_\textrm{ren}$, $\langle\hat{T}_{uu}^-\rangle^\subU_\textrm{ren}$ and $\langle\hat{T}_{vv}^-\rangle^\subU_\textrm{ren}$
as functions of the scalar field mass $\mu$, for 
coupling $\xi=0$ and BH charge $Q=0.8M$.
}
\label{fig:Tyy_CH_Mass}
\end{figure}

In the  case of massless fields, it can be shown~\cite{McNamara1978,Kehle2019,PhysRevD.20.1260,MassiveRNPhiSq} that 
$q_{l}$,
$h_l(0)$
and
the coefficient of a subleading  $\log(\log|f|)$ 
term 
all
vanish, 
so that the VP is regular as $r\to r_{-}$.
In its turn, in the massive case, it is $q_{l}\neq 0$
\cite{McNamara1978,Kehle2019}
and in  Fig.~\ref{fig:log_coeff_num} we plot our
 numerical computation of 
 $k(\mu,Q)$ using
 Eq.~\eqref{eq:nearhorizonclosed} (see 
 Sec.~V
 in SM for numerical details).
 This figure  shows that $k(\mu\neq 0,Q)$ 
 is dominated by the $l=0$ mode and that it
 is (generically) nonzero. 
 This establishes that, for  fields with finite mass, $\langle \hat{\Phi}^{2}\rangle_{\textrm{ren}}^{\subH/\subU}$ diverges logarithmically at the CH as per \eqref{eq:asympt VP}. 
 In the large-mass limit, however, it can be seen 
 (Fig.~1 in SM) 
 that $k(\mu,Q)$ vanishes asymptotically.
 
\paragraph*{Results for fluxes.}

We now turn to the RSET, which, as opposed to the VP, also depends on the coupling parameter  $\xi$.
Of particular interest are its energy flux components on the CH (denoted by a superscript $^-$): $\langle \hat{T}_{yy}^-\rangle_{\textrm{ren}}^\subA$, with $y=u$ or $v$.
As Ref.~\cite{Zilberman:2019buh} shows, 
the fluxes have a notable accumulation effect on the form of the metric  when taking into account backreaction  via the semiclassical Einstein equations.
In particular, 
an observer infalling towards the CH will experience rapid contraction if $\langle \hat{T}_{vv}^-\rangle_{\textrm{ren}}^\subA>0$ and rapid expansion if $\langle \hat{T}_{vv}^-\rangle_{\textrm{ren}}^\subA<0$.
The high rapidity can be seen by transforming 
from Eddington-Finkelstein
to
the {\it regular}
Kruskal coordinates: it is 
$\langle T_{VV}^-\rangle^\subA_{\textrm{ren}}=
\frac{1}{\kappa_-^2 V^2}
\langle T_{vv}^-\rangle_\textrm{ren}^\subA$.
This means that, if  
$\langle \hat{T}_{vv}^-\rangle_{\textrm{ren}}^\subA$  
is nonzero, then the  
flux
$\langle T_{VV}^-\rangle^\subA_{\textrm{ren}}$ 
diverges  
quadratically
at  the
CH. 
In its turn, a positive/negative $\langle T_{uu}^-\rangle^\subA_{\textrm{ren}}$ may lead to a 
contraction/expansion of the CH itself.

Let us proceed with the calculation of the RSET.
It was shown in~\cite{taylor2022mode} that the RSET 
for a spherically-symmetric BH
can be calculated in a simple manner from the Green function and its derivatives (see Eq\MCt{s}.~(13) and (16) in Ref.~\cite{taylor2022mode}).
We then carried out near-CH asymptotics of the RSET in a manner similar  to the way we  
did
it for the VP  in the previous section.
Our result for the angular components $\langle \hat{T}^{\theta}{}{}_{\theta}\rangle_{\textrm{ren}}^{\subA}=\langle \hat{T}^{\phi}{}{}_{\phi}\rangle_{\textrm{ren}}^{\subA}$ on the CH is the following \cite{RSETLong}. For $\xi\neq 1/4$ and in at least one of the two states, they diverge at least as strongly as $\cos(2\mu r_*)/(f r_*)$ (aside constants);
so they oscillate and diverge more strongly than for massless fields, in which case they diverge like $1/(fr_*^{n+2})$, with $n=2$ for  H-H   and $n=3$ for Unruh when $\xi\neq 1/4$ (e.g., see~\cite{Lanir:2018vgb} for the trace for $\xi=0$).
For $\xi=1/4$, we obtain for massive fields:
 \begin{align}
\label{eq:Tab_IH}
    \langle \hat{T}^{\theta}{}{}_{\theta}\rangle_{\textrm{ren}}^{\subA}=\langle \hat{T}^{\phi}{}{}_{\phi}\rangle_{\textrm{ren}}^{\subA} & \sim \mathcal{L}\,\log|f|,&r\to r_-, \quad \xi=1/4,
\end{align}
where
\begin{equation}
\mathcal{L}:=\sum_{l=0}^\infty \frac{l+1/2}{r_-^2}\left(l(l+1)+\frac{r_+}{2r_-}\right)\,h_l(\mu).
\end{equation}
Thus, while the angular components of the RSET on the CH for $\xi=1/4$ are finite for massless fields, they (generically) diverge logarithmically for massive fields. As for the fluxes on the CH, we obtain
\begin{align}
\label{eq:Tuu_massive_H}
        \langle \hat{T}_{uu}^-\rangle_{\textrm{ren}}^\subH=\langle \hat{T}_{vv}^-\rangle_{\textrm{ren}}^\subH  = \mathcal{K^\subH},
        \quad
            \langle \hat{T}_{yy}^-\rangle_{\textrm{ren}}^\subU=  \mathcal{K
            ^\subH
            }+\mathcal{J}_y,
\end{align}
where 
\begin{align}
\label{eq:KH}
    \mathcal{K
    ^\subH
    }&:=\frac{1}{8\pi^2}\sum_{l=0}^{\infty}(2l+1)\Bigg(\mathcal{Y}
    _{l}+8\pi^2\xi \,\kappa_-^2\,h_l(\mu)\Bigg),
\end{align}
with 
\begin{align}
\label{eq:Yl}
   & \mathcal{Y}_l
    :=\int_{0}^{\infty}\frac{d\omega}{r_-^2} \,\omega\Bigg[\coth\left(\frac{\pi\omega}{\kappa_{+}}\right)|A_{\omega l}|^{2}+\nonumber\\
   &
    \csch\left(\frac{\pi\omega}{\kappa_{+}}\right)\Re\{ A^{\textrm{up}}_{\omega l}(A_{\omega l}B_{\omega l})\}\Bigg]+\frac{\kappa_+^2-\kappa_-^2}{24r_-^2},
\end{align}
\begin{align}
\label{eq:Jy}
&\mathcal{J}_
y
:= \frac{1}{32\pi^{2}r_-^2}\sum_{l=0}^{\infty}(2l+1)
\cdot 
\nonumber\\ &
\int_\mu^\infty\!\! d\omega\, \omega\left(1-\coth\left(\frac{\pi\omega}{\kappa_+}\right)\right)(1-|A_{\omega l}^\textrm{up}|^2)\left(|A_{\omega l}|^2+\delta^v_y\right).
\end{align}
Eqs.~\eqref{eq:Tuu_massive_H}-\eqref{eq:Jy} show that the coefficient of the leading divergence of the fluxes is independent of $\xi$ in the massless case (since $h_l(0)=0$), whereas it depends on $\xi$ in the massive case.

\begin{figure}
\centering
\includegraphics[width=\linewidth]{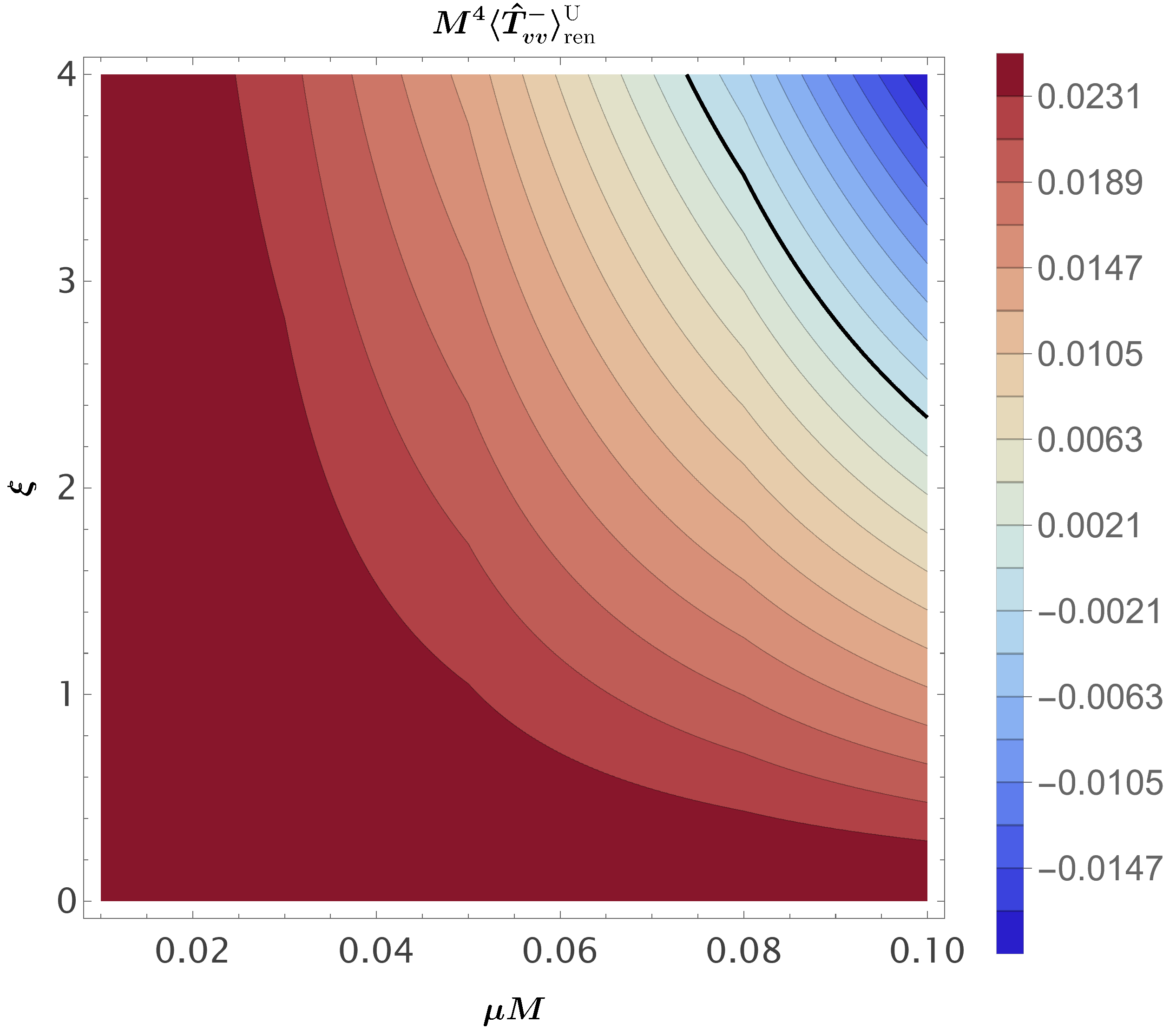}
\caption
{CH flux $\langle\hat{T}_{vv}^-\rangle^\subU_\textrm{ren}$ as a function of the field mass $\mu M\in[0,0.1]$ and coupling $\xi\in[0,4]$ for BH charge $Q=0.8M$.
The thick black line corresponds to $\langle\hat{T}_{vv}^-\rangle^\subU_\textrm{ren}=0$.
}  
\label{fig:ContourPlotTuuHH}
\end{figure}

\begin{figure}
\centering
\includegraphics[width=\linewidth]{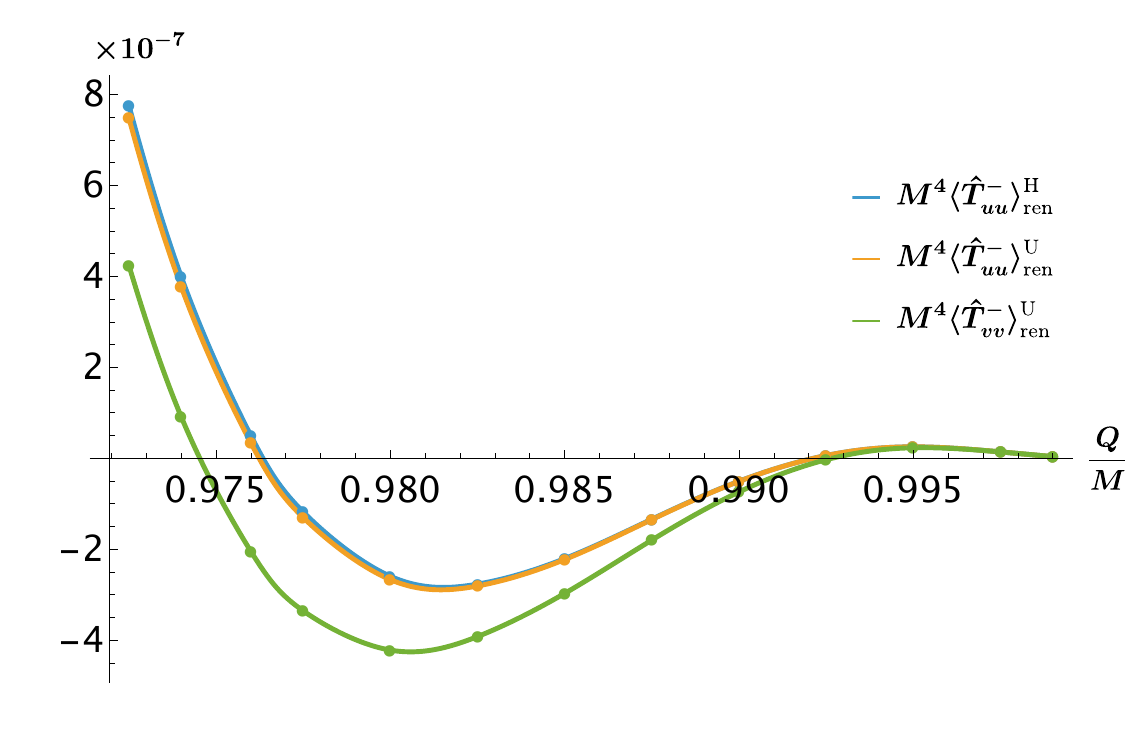}
\caption{CH fluxes $\langle\hat{T}_{uu}^-\rangle^\subH_\textrm{ren}=\langle\hat{T}_{vv}^-\rangle^\subH_\textrm{ren}$, $\langle\hat{T}_{uu}^-\rangle^\subU_\textrm{ren}$ and $\langle\hat{T}_{vv}^-\rangle^\subU_\textrm{ren}$ as functions of the BH charge $Q$ for coupling $\xi=0$ and field mass $\mu M=0.1$.
}
\label{fig:Tyy_CH_Q}
\end{figure}

We  
used 
Eq.~\eqref{eq:Tuu_massive_H}
to calculate and plot the fluxes on the CH.
In Fig.~\ref{fig:Tyy_CH_Mass}, we plot them as a function of the field mass $\mu$.  
The numerical results for a massless field are in agreement with Fig.~1 of the SM in Ref.~\cite{Zilberman:2019buh}.  
Although the magnitude in the H-H state is larger than in the Unruh state, within the range of investigation,
the magnitude of the fluxes increases with the mass, with the rate of increase being stronger in  Unruh. Fig.~\ref{fig:ContourPlotTuuHH} presents a contour plot of  $\langle \hat T_{vv}^-\rangle_\textrm{ren}^\subU$ as a function of the field mass $\mu$ and coupling $\xi$; the corresponding plots for  $\langle \hat T_{uu}^-\rangle_\textrm{ren}^\subU$  and for the fluxes in the H-H state are qualitatively the same. This figure shows that there are certain fine-tuned values of $\mu$ and $\xi$ for which the flux at the CH is zero.

\begin{figure}
\centering
\includegraphics[width=\linewidth]{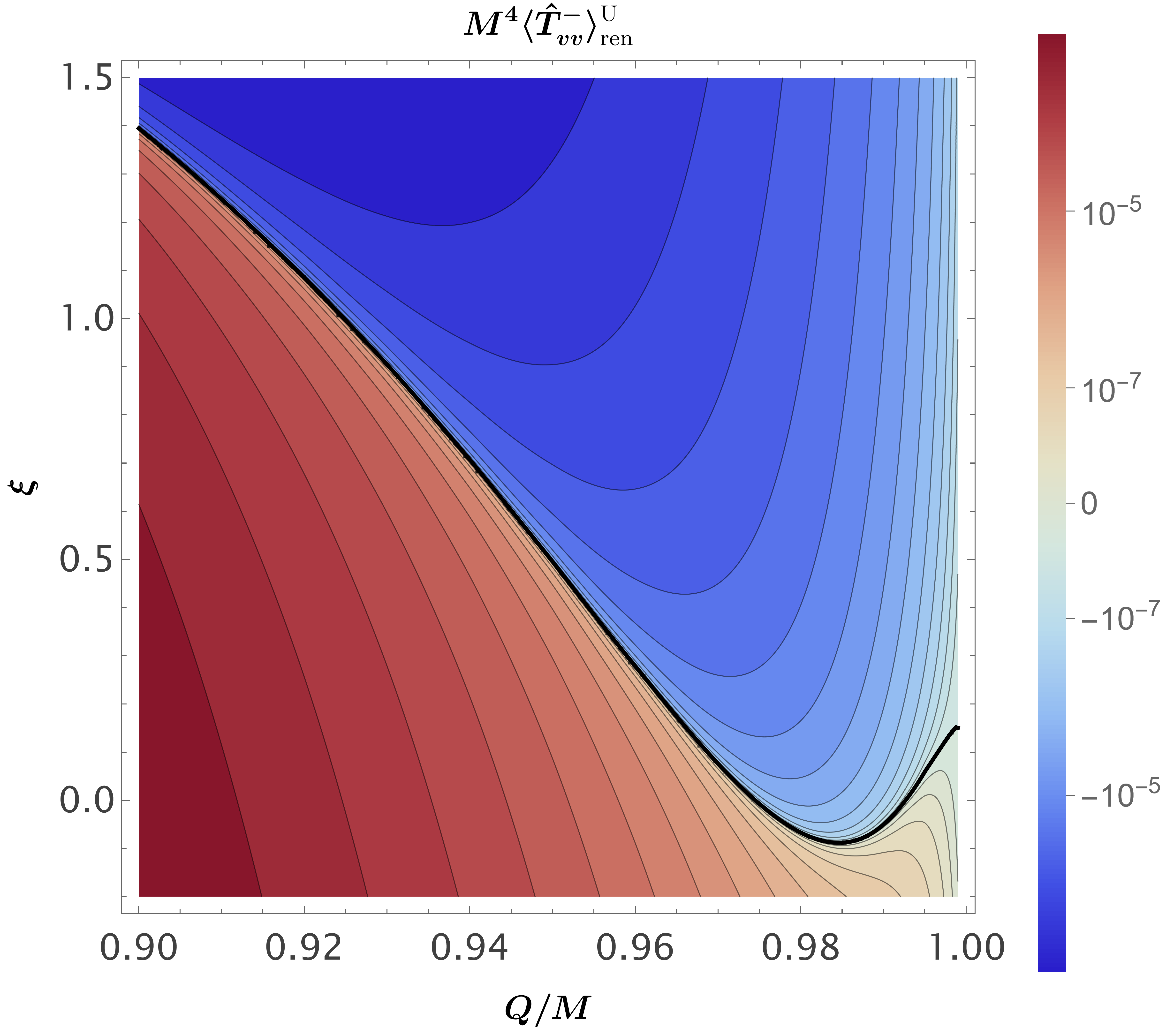}
\caption{
CH flux $\langle\hat{T}_{vv}^-\rangle^\subU_\textrm{ren}$ as a function of  the BH charge $Q/M\in[0.9,0.999]$ and  coupling $\xi\in[-0.2,1.5]$ for field mass $\mu M=0.1$. The thick black line corresponds to $\langle\hat{T}_{vv}^-\rangle^\subU_\textrm{ren}=0$ exactly.
}
\label{fig:Contour-xi-Q}
\end{figure}

We also produced analogous plots to investigate the dependence of the CH fluxes on the BH charge $Q$, with a particular focus on the near-extremal regime. Figs.~\ref{fig:Tyy_CH_Q} and \ref{fig:Contour-xi-Q} show the results for a  field with mass $\mu M=0.1$. The first figure indicates that the  fluxes take on similar values in both quantum states and vanish in the extremal limit. In addition, comparing Fig.~\ref{fig:Tyy_CH_Q} with the corresponding figure for a massless field (see Fig.~1 in Ref.~\cite{Zilberman:2019buh}), it can be noticed that the fluxes $\langle \hat T_{yy}^-\rangle_\textrm{ren}^\subA$ for a field with mass $\mu M=0.1$ change sign twice instead of once in the massless case. Specifically, for $Q/M\gtrsim0.9925$, the fluxes are positive
(leading to 
contraction for an infalling observer
or CH for, respectively, $y=v$ or $y=u$)
for a massive field while remaining negative 
(leading, instead, to  
expansion)
for a massless field. 
This result is particularly surprising, since one would naively expect that, in the extremal limit $r_+- r_-\to 0$, the sign of the influx on the CH is the same as on the event horizon, where it is positive  (as corresponding to Hawking radiation) -- see the argument in~\cite{2021PhRvD.104b4066Z} for massless fields.
In its turn, the contour plot in Fig.~\ref{fig:Contour-xi-Q} of $\langle \hat T_{yy}^-\rangle_\textrm{ren}^\subA$ as a function of the BH charge $Q$ and coupling $\xi$ further shows that, for subextremal $Q<M$, there exist fine-tuned values of  
$\xi$ for which the fluxes vanish; 
in the extremal limit ($Q\to M^-$), the fluxes seem to vanish
(within our numerical precision) for all $\xi$.
As before, the contour plots for the components in the H-H state are qualitatively the same as those in the Unruh state.

\paragraph*{Conclusions.}

In this Letter we have investigated the behaviour of quantum {\it massive} fields in the interior of a RN BH. Whereas there has  
been 
some
recent
results for similar investigations in the case of massless fields in various BH spacetimes,
as well as for the influx for massive fields on the CH of RNdS, we have presented, to the best of our knowledge, the first results for a quantum massive field on the CH of 
an asymptotically-flat BH.

The obtainment of our results for a massive field has been made possible by implementing, for the first time, the extended coordinate method directly (i.e., without Euclideanization) to a Lorentzian setting.
This method is, not only very efficient, but also, and as opposed to other existing renormalization methods,  quite agnostic to the particular parameters (such as the mass) of the quantum field.

Our results show a very significant difference between massive and massless fields for the VP $\langle\hat\Phi^2\rangle_{\mathrm{ren}}^\subA$ 
and the angular components $ \langle \hat{T}^{\theta}{}{}_{\theta}\rangle_{\textrm{ren}}^{\subA}=\langle \hat{T}^{\phi}{}{}_{\phi}\rangle_{\textrm{ren}}^{\subA}$  of the RSET for $\xi=1/4$
on the CH: whereas they are
finite in the massless case, they
(generically) diverge in the massive case.
For $\xi\neq 1/4$, the angular components diverge more strongly for massive fields in at least one of the two states.
This suggests a meaningful physical distinction between long range (massless) and short range (massive) quantum fluctuations near the CH.

As for the fluxes, 
we identified regions of parameter space where $ \langle T_{yy}^-\rangle_\textrm{ren}^\subA$ is positive/negative, leading to a contraction/expansion of an infalling observer ($y=v$) or CH itself ($y=u$). The 
regions of positive/negative signs differ from  those for a massless field
 and come with a surprising sign twist near extremality, 
 where the signs of  the influx on the CH and on the event horizon differ -- this might be 
 understood
 from the fact that massive fields are inherently more localized. 
Importantly, our results show that, except at extremality  or for fine-tuned values of the field and subextremal BH parameters, $\langle T_{vv}^-\rangle_\textrm{ren}^\subA$ is nonzero, and thus 
$\langle T_{ VV}^-\rangle^\subA_{\textrm{ren}}$ diverges like $V^{-2}$.
Notably, this divergence is stronger  than in the classical scenario, 
where the metric is continuous on the CH~\cite{kehle2024strong,2018CMaPh.360..103V,1998PhRvD..58b4018H,PhysRevD.63.064032,PhysRevD.88.024054,2004PhRvD..70d4018B,PhysRevD.68.044013}.  
As argued in \cite{hollands2020quantum} (in the context of RNdS), such 
$V^{-2}$ divergence should suffice to convert the CH into a strong singularity, when considering a fixed background. 
While this is a similar divergence to that in the massless case~\cite{Zilberman:2019buh},
the massive case presents more oscillations as a function of the BH charge $Q$ near extremality, thus affecting the regions of values of $Q$ where an infalling observer would experience rapid contraction ($ \langle T_{vv}^-\rangle_\textrm{ren}^\subA>0$) vs rapid expansion ($ \langle T_{vv}^-\rangle_\textrm{ren}^\subA<0$). 
That is, our
results suggest that introducing a nonzero mass for the scalar field changes the nature of the tidal deformations in BHs with values of $Q$ close to the extremal limit. 
In the actual extremal BH limit, it is $ \langle T_{vv}^-\rangle_\textrm{ren}^\subA\to 0$, so that the conclusion of  
a quadratic
divergence of the Kruskal flux $ \langle T_{VV}^-\rangle_\textrm{ren}^\subA$ no longer necessarily follows.
The extremal case, where massless classical perturbations 
possess {\it finite} local energy on the CH~\cite{Gajic:2015csa,murata2013happens},
deserves a separate investigation which we leave for future work.

Relatedly,
and interestingly,
Ref.~\cite{Klein2024long} showed that, in the case of a massless field in RNdS,  
correlations of the ingoing Unruh flux
on the CH
are not negligible over macroscopic distances (angles), that they
are of the same order of magnitude as the square of the flux, and that they diverge in a Hadamard-state-independent manner like $V^{-4}$;
this signals
a breakdown of the semiclassical approximation.
Their results are based on the mode $l=0$ being the dominant contribution to the fluxes. Since we have found a similar $l=0$-dominance  for massive fields in RN, this suggests that  quantum fluctuations on the CH might also  be non-negligible over macroscopic distances in our setting. However, a thorough, proper analysis of  quantum  
correlations and their backreaction effects
is beyond the scope of this work.

Last but not least, we would like to extend our results to the more astrophycally relevant case of a Kerr BH. Unfortunately, however, the state-of-the-art of the various renormalization methods does not yet allow for their immediate implementation for massive fields in Kerr (state-subtraction~\cite{christensen1977trace,candelas1980vacuum,hollands2020quantum,2022PhRvL.129z1102Z,McMaken:2024fvq,PhysRevLett.132.121501} might work {\it exactly} on the CH but not everywhere else in the interior). Thus, a next step should be the development of such a method.


\section*{Acknowledgments}
We are thankful to Stefan Hollands, Jochen Zahn and Noa Zilberman for
useful discussions. L.P. acknowledges funding from Taighde Éireann - Research Ireland under Grant number GOIPG/2024/4003.

\bibliography{biblio_PRL}
\flushcolsend

\clearpage
\newpage
\pagestyle{empty}
\null
  \AddToShipoutPictureBG*{%
    \AtPageLowerLeft{%
      \includegraphics[page=1, width=\paperwidth, height=\paperheight]{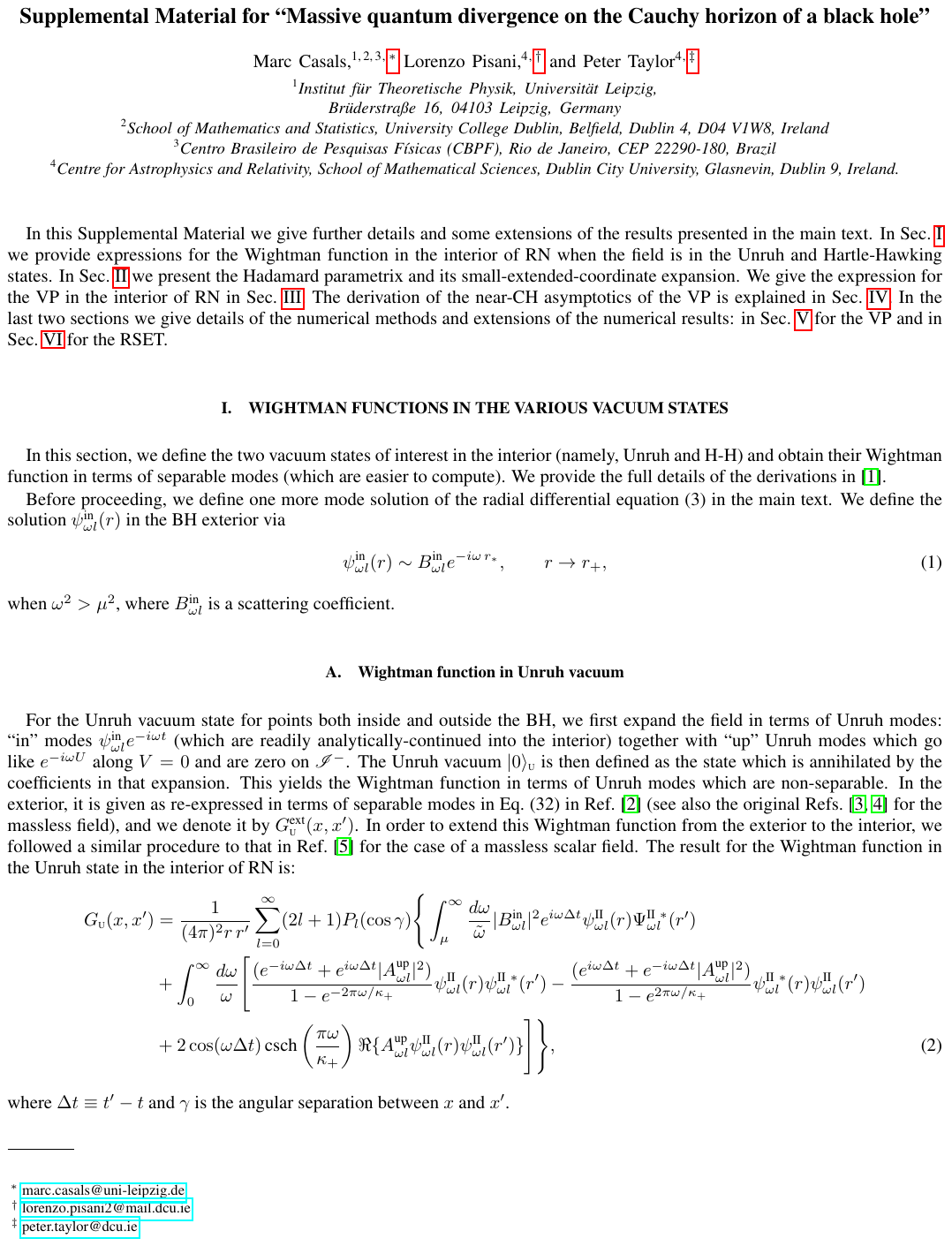}%
    }%
  }%
  
 \clearpage
 \newpage
 \pagestyle{empty}
 \null
   \AddToShipoutPictureBG*{%
     \AtPageLowerLeft{%
       \includegraphics[page=2, width=\paperwidth, height=\paperheight]{SupplementalMaterial.pdf}\newpage%
     }%
   }%
   
   \clearpage
   \newpage
   \pagestyle{empty}
   \null
     \AddToShipoutPictureBG*{%
       \AtPageLowerLeft{%
         \includegraphics[page=3, width=\paperwidth, height=\paperheight]{SupplementalMaterial.pdf}\newpage%
       }%
     }%
	 \clearpage
	 \newpage
	 \pagestyle{empty}
	 \null
	   \AddToShipoutPictureBG*{%
	     \AtPageLowerLeft{%
	       \includegraphics[page=4, width=\paperwidth, height=\paperheight]{SupplementalMaterial.pdf}\newpage%
	     }%
	   }%
	   \clearpage
	   \newpage
	   \pagestyle{empty}
	   \null
	     \AddToShipoutPictureBG*{%
	       \AtPageLowerLeft{%
	         \includegraphics[page=5, width=\paperwidth, height=\paperheight]{SupplementalMaterial.pdf}\newpage%
	       }%
	     }%
		 \clearpage
		 \newpage
		 \pagestyle{empty}
		 \null
		   \AddToShipoutPictureBG*{%
		     \AtPageLowerLeft{%
		       \includegraphics[page=6, width=\paperwidth, height=\paperheight]{SupplementalMaterial.pdf}\newpage%
		     }%
		   }%

\end{document}